\documentclass[referee]{aa}
\usepackage{supertabular}
\usepackage{epsf} 
 
\def\tempest%
{\begin{array}{ccc} 
1 & 1 & 1 \\ 
1 & 1 & 1 \\ 
4 & 3 & 8 
\end{array}}

\def\kms{{\rm km}\,{\rm s}^{-1}}

\def\bu{{\bf u}}
\begin{document}

\title{Photometric Constraints on Microlens Spectroscopy of EROS-BLG-2000-5}
\author 
{
C.~Afonso\inst{1},
J.N.~Albert\inst{2},
J.~Andersen\inst{6},
R.~Ansari\inst{2}, 
\'E.~Aubourg\inst{1}, 
P.~Bareyre\inst{1,4}, 
F.~Bauer\inst{1},
G.~Blanc\inst{1},
A.~Bouquet\inst{4},
S.~Char$^{\dag}$\inst{7},
X.~Charlot\inst{1},
F.~Couchot\inst{2}, 
C.~Coutures\inst{1}, 
F.~Derue\inst{2}, 
R.~Ferlet\inst{3},
P.~Fouqu\'e\inst{9,10},
J.F.~Glicenstein\inst{1},
B.~Goldman\inst{1},
A.~Gould\inst{8,4},
D.~Graff\,\inst{8,1},
M.~Gros\inst{1}, 
J.~Ha\"{\i}ssinski\inst{2}, 
J.C.~Hamilton\inst{4},
D.~Hardin\inst{1},
J.~de Kat\inst{1}, 
A.~Kim\inst{4},
T.~Lasserre\inst{1},
L.~LeGuillou\inst{1},
\'E.~Lesquoy\inst{1,3},
C.~Loup\inst{3},
C.~Magneville \inst{1}, 
B.~Mansoux\inst{2}, 
J.B.~Marquette\inst{3},
\'E.~Maurice\inst{5}, 
A.~Milsztajn \inst{1},  
M.~Moniez\inst{2},
N.~Palanque-Delabrouille\inst{1}, 
O.~Perdereau\inst{2},
L.~Pr\'evot\inst{5}, 
N.~Regnault\inst{2},
J.~Rich\inst{1}, 
M.~Spiro\inst{1},
A.~Vidal-Madjar\inst{3},
L.~Vigroux\inst{1},
S.~Zylberajch\inst{1}
\\   \indent   \indent
The EROS collaboration
}
%
\institute{
CEA, DSM, DAPNIA,
Centre d'\'Etudes de Saclay, F-91191 Gif-sur-Yvette Cedex, France
\and
Laboratoire de l'Acc\'{e}l\'{e}rateur Lin\'{e}aire,
IN2P3 CNRS, Universit\'e de Paris-Sud, F-91405 Orsay Cedex, France
\and
Institut d'Astrophysique de Paris, INSU CNRS,
98~bis Boulevard Arago, F-75014 Paris, France
\and
Coll\`ege de France, Physique Corpusculaire et Cosmologie, IN2P3 CNRS, 
11 pl. M. Berthelot, F-75231 Paris Cedex, France
\and
Observatoire de Marseille,
2 pl. Le Verrier, F-13248 Marseille Cedex 04, France
\and
Astronomical Observatory, Copenhagen University, Juliane Maries Vej 30, 
DK-2100 Copenhagen, Denmark
\and
Universidad de la Serena, Facultad de Ciencias, Departamento de Fisica,
Casilla 554, La Serena, Chile
\and
Departments of Astronomy and Physics, Ohio State University, Columbus, 
OH 43210, U.S.A.
\and
Observatoire de Paris-Meudon DESPA, F92195 Meudon CEDEX, France
\and
European Southern Observatory, Casilla 19001, Santiago 19, Chile
}

\mail{gould@astronomy.ohio-state.edu,\ glicens@hep.saclay.cea.fr}
\titlerunning{Contraints on EROS-BLG-2000-5 Spectroscopy}
\authorrunning{EROS Collaboration}   
\maketitle
\begin{abstract} 

	We apply EROS photometric data to interpret previously published
Keck and VLT spectra of the binary-microlens caustic-crossing event
EROS-BLG-2000-5.  We show that the VLT data imply that the outer $\sim 4\%$
of the limb of the K3-giant source is strongly in emission in H$\alpha$,
in contradiction to available models of the photosphere.  This conflict
could be resolved if the integrated H$\alpha$ emission from the chromosphere
were equal to 8\% of the integrated H$\alpha$ absorption from the source
as a whole.
These conclusions regarding the extreme limb are almost
completely model-independent.  We also present a general method for using
the photometric data to facilitate direct comparison between the 
atmospheric model and the spectroscopic data.   While this method 
has some model-dependent features, it is fairly robust and can serve
to guide the analysis of spectra while more detailed models of the
lens geometry are being developed.  In addition, we find that
the color of the limb of the source (outer 5.5\% by radius) is 
$\Delta(V-I)\sim 0.37$
redder than the source as a whole, so that it has the color of a M0 giant.
\end{abstract}

\thesaurus{}

\section{Introduction} 

	Microlensing provides a potentially powerful probe of stellar
atmospheres.  If a lens caustic (region of formally infinite magnification)
passes over the face of the source, then different parts of the atmosphere
become highly magnified at different times.  By combining a time series
of photometric or spectroscopic observations, one can therefore hope to 
deconvolve the spatial structure of the atmosphere.  For point lenses,
the caustics are point-like, and therefore the probability of such
a caustic crossing is small.  However, if the lens is a binary with 
components separated by of order the Einstein radius, the caustics form
one to three concave polygons whose total length is of order the
Einstein radius, and therefore
the probability of a caustic crossing is much larger.

	Intensive photometric observations of four such binary caustic 
crossing events have yielded limb-darkening measurements 
(Afonso et al.\ 2000; Albrow et al.\ 1999,2000,2001a).  A more ambitious 
project would be to obtain a similar intensive series of spectroscopic 
measurements.  Since light from the limb of the star originates higher
in the atmosphere than light from the center, deconvolution of a
set of spectral measurements effectively resolves the atmosphere as a 
function of height (Valls-Gabaud 1998; Heyrovsk\'y, Sasselov \& Loeb 2001;
Gaudi \& Gould 1999).  Until the advent of microlensing, such spatially
resolved spectra had been obtained only for the Sun.

	Early spectra of caustic-crossing events were obtained by
Lennon et al.\ (1996) and Alcock et al.\ (1997), but no clear differences
were detected between spectra taken at different times.  

	The binary caustic crossing event EROS-BLG-2000-5 has now
yielded the first big breakthrough.  This event was alerted by 
EROS\footnote{http://www-dapnia.cea.fr/Spp/Experiences/EROS/alertes.html}
on 5 May 2000.  On 8 June 2000, 
MPS\footnote{http://bustard.phys.nd.edu/MPS/index.html} issued an anomaly 
alert, saying that a caustic crossing was in progress.  Subsequent intensive
observations by PLANET\footnote{http://thales.astro.rug.nl/~planet/}
allowed them to predict the time of the second caustic
crossing and, very importantly, that this crossing would last an 
unusually long 4 days.  This prediction permitted
two groups to acquire spectra with large telescopes on several successive
nights:  Castro et al.\ (2001) obtained HIRES ($R\sim 40,000$)
spectra using Keck on the last two nights of the crossing, while
Albrow et al.\ (2001b) obtained low-resolution ($R\sim 1000$) FORS1 spectra
using VLT on all four nights.  
In addition, Albrow et al.\ (2001b) obtained a spectrum well
before the caustic crossing, when the magnification of the source was
approximately uniform.

To date, each group has published only the H$\alpha$ lines,
but both sets of spectra cover several thousand \AA, and could potentially
yield a wealth of information about the K3 giant source.

	Here we use EROS photometric observations of this event to aid
in the interpretation of the spectral data that have been published.
In \S\ 2, we present the EROS photometric data of the event.
In \S\ 3, we present an essentially model-independent argument
to show that the VLT data imply that the outer $\sim 4\%$ of the limb of 
the source is very strongly in emission in H$\alpha$.  In \S\ 4, we develop 
a general method to apply photometric data of a caustic crossing to
facilitate comparison between spectroscopic data and atmospheric models
and apply this method to EROS-BLG-2000-5. In \S\ 5, we characterize
the limits of our approach.  In \S\ 6, we measure the color of
the source limb by applying a slightly modified version of our approach
to the EROS photometric data.  Finally, in \S\ 7, we speculate on the
possibility that the H$\alpha$ emission detected from the limb of the 
source is actually due to the chromosphere, and quantify the strength
of chromospheric emission required to explain the observations.

	If the geometry of EROS-BLG-2000-5 were perfectly modelled, then
there would be no need for the techniques introduced in this paper.
However, the event is quite complex and difficult to model, so additional
techniques are required to begin quantitative investigation of the
spectroscopic data while a better geometric model is being developed.
Moreover, it is likely that in future events, one will face a qualitatively
similar situation, and the methods presented in this paper will be
applicable long before the event as a whole is analyzed at a satisfactory
level.

\section{Photometric Data}

	EROS observations were carried out using the 1 m Marly telescope
at La Silla, Chile, in two bands $V_E$ and $I_E$.  EROS $V_E$ is centered
midway between Johnson $V$ and Cousins $R$, while
EROS $I_E$ is similar to Cousins $I$, but broader.
The transformation between the Eros system ($V_E$, $I_E$) and the
standard Johnson-Cousins ($V$,$I$) system was determined  by observing
Landolt (1992)  standards near the galactic poles  
and Paczynski et al (1999)  secondary standards in the Baade window:
\begin{equation}
V_E = 0.71\,V + 0.29\,I +\,{\rm const},\quad
I_E = -0.02\,V + 1.02\,I +\,{\rm const},
\label{eqn:vitrans}
\end{equation} 
These equations taken together together imply
$\Delta(V_E-I_E) = 0.73\Delta(V-I).$  
The first of equations (\ref{eqn:vitrans}) implies that the
centroid of the $V_E$ band is approximately 6300 \AA.
We therefore present mainly $V_E$ data because its centroid is
close to H$\alpha$.
EROS observations are normally carried out in survey mode,
but owing to the importance of this event we devoted all available
time to it during the caustic crossing.  Generally, the weather was good,
but on the final night there were intermittent clouds, and then finally
we were clouded out entirely four hours before dawn.

The data were initially
reduced on site using our standard PEIDA (Ansari, et al.\ 1996) 
PSF-based software, but were then
reprocessed using the ISIS (Alard, 2000) image subtraction program.  We
then determined the zero point of the image-subtracted photometry by
fitting for the mean offset between it and the PEIDA photometry.
Figure \ref{fig:one} shows the naive magnification 
$A_V$, which we determine 
by dividing the measured flux by the baseline flux.  The curve will be
explained in \S\ 4, but from the data points alone, it is clear that
the caustic crossing ended some time before JD$'=1733.6506$:  the light
curve is almost perfectly flat commencing at this time and continuing 
for 58 minutes, with a measured slope (combining information from both
bands) of $d\ln A/dt = 0.11\pm0.08\rm\,day^{-1}$.  (Here 
JD$'=$JD$-2450000$.)

The crosses indicate the times of the Keck and VLT observations.  Note that
the VLT observations were all coincident with our photometric observations,
while the Keck observations both took place about 2.2 hrs after the 
end of our night.

\section{H$\alpha$ in Emission?}

The last VLT equivalent width (EW) measurement, which was taken
 just before the source
finally exited the caustic, is much lower than any of the other EW
measurements.  As we will show a few paragraphs below, 
this can be interpreted as indicating strong H$\alpha$
emission from the edge of the star.  In
order to quantify this conclusion, we must first derive some relations
between the photometric and spectroscopic data.

	Let $A(\bu)$ be the magnification field of the binary as
a function of angular position $\bf u$ normalized to the Einstein radius, 
let $E(\bu)$ be the equivalent width of the H$\alpha$ line, and let
$S({\bf u})$ be the source surface brightness in the neighborhood of this
line.  We will split the magnification field into two components
$A(\bu)=A_2(\bu) + A_3(\bu)$ corresponding respectively to the two 
images whose magnification diverges (and then vanishes) at the caustic,
and the three non-divergent images.  The observed magnification is
then,
\begin{equation}
\overline{A}(\bu_c) = {\int d^2 u A(\bu) S(\bu)\over\int d^2 u S(\bu)}
= A_3(\bu_c) + \overline{A_2}(\bu_c),
\label{eqn:abar}
\end{equation}
where
\begin{equation}
\overline{A_2}(\bu_c) \equiv
{\int d^2 u A_2(\bu) S(\bu)\over F_s},\qquad 
F_s \equiv \int d^2 u S(\bu),
\label{eqn:abar2}
\end{equation}
and where we have assumed that the non-divergent magnification field can
be replaced by its value at the center of the source $\bu_c$.  This
is strictly true whenever this magnification field is either flat or
has a uniform slope.  We will discuss in \S\ 5 why we believe this to be an 
extremely good approximation in the present case.

	Similarly, the observed H$\alpha$ EW will be
\begin{equation}
\overline{E}(\bu_c) 
= {\int d^2 u A(\bu)S(\bu)E(\bu)\over\int d^2 u A(\bu)S(\bu)}
= {A_3(\bu_c)E_0 + \overline{A_2}(\bu_c)\overline{E_2}(\bu_c)
\over A_3(\bu_c) + \overline{A_2}(\bu_c)},
\label{eqn:hbar}
\end{equation}
where, $E_0$ is the EW of the unmagnified source, and
\begin{equation}
\overline{E_2}(\bu_c) \equiv 
{\int d^2 u A_2(\bu) S(\bu)E(\bu)\over F_s\overline{A_2}(\bu_c)}
\label{eqn:hbar2}
\end{equation}
is the EW that would be seen if a separate spectrum could be taken
of the two highly magnified images.  Thus, if $E_0$ and $\overline{E}(\bu_c)$
are measured spectroscopically and if the fraction of light coming
from the highly magnified images 
$f(\bu_c)\equiv {\overline A_2}(\bu_c)/{\overline A}(\bu_c)$
is known from the photometric light curve, then one can invert equation
(\ref{eqn:hbar}) to determine $\overline{E_2}(\bf u_c)$,
\begin{equation}
{\overline{E_2}\over E_0} = 1 + {\Delta\over f},\quad
\Delta \equiv {\overline{E} - E_0\over E_0},\quad
f \equiv {\overline{A_2}\over \overline{A}}.
\label{eqn:h2eval}
\end{equation}

\begin{figure}
\epsfxsize \hsize
\epsfbox{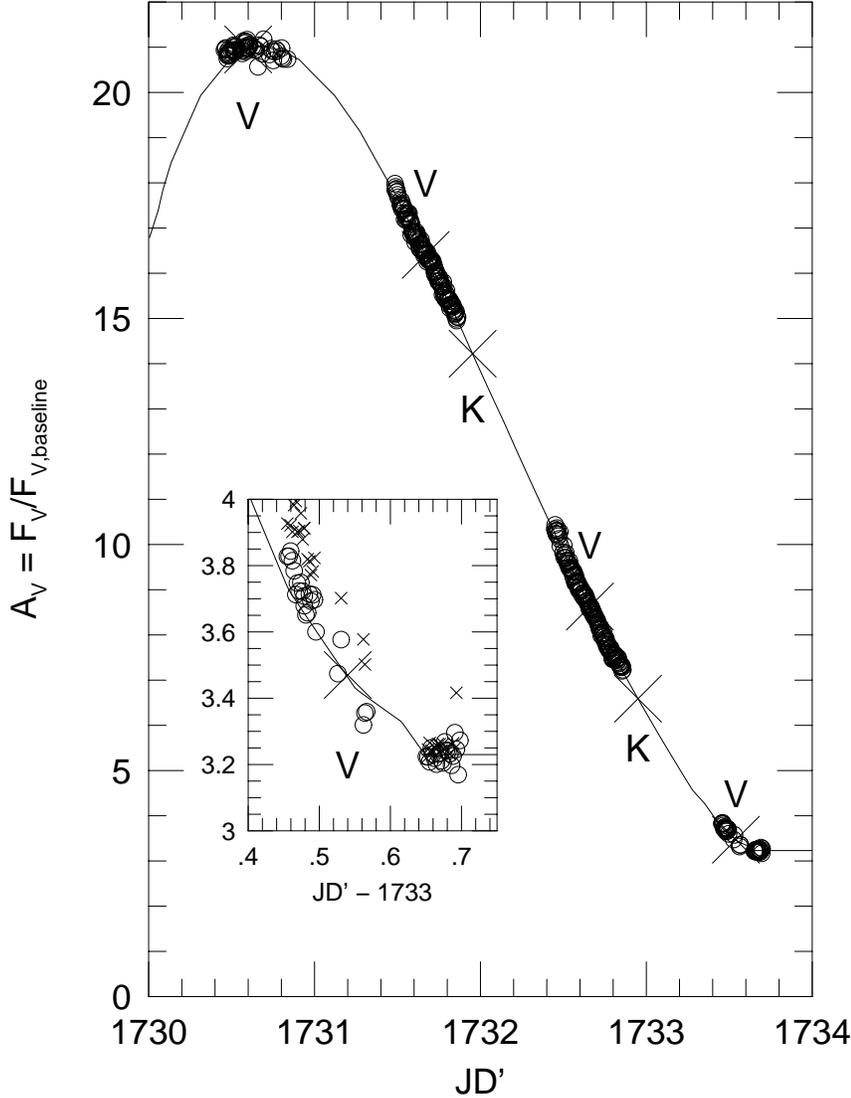}
\caption[junk]{\label{fig:one}
Magnification $A_V$ of a K3 giant during the second caustic crossing
of EROS-BLG-2000-5.  Open circles show the fluxes in the EROS $V_E$ band
divided by the baseline flux (i.e., assuming no blending.)  Large
crosses represent the times of spectroscopic observations by
Albrow et al.\ (2001b) at the VLT (V) and by Castro et al.\ (2001) at
Keck (K).  The curve is the projection of the simple quadratic model
shown in Fig.\ \ref{fig:two} into (magnification,time) coordinates
using the stellar-profile function $G(\eta)$ [eq.\ (\ref{eqn:geval})]
derived from Fig.\ 1 of Albrow et al.\ (2001b).
(Irregularities are due to the difficulty of transcribing the figure.)\ \
The insert also shows EROS $I_E$ points (crosses) for the last night.  
 From the two bands together, the slope of the magnification during the
hour after the caustic exit at JD$'=1733.65$ is $d \ln A/d t=
(0.11\pm 0.08)\,\rm day^{-1}$.  Here, JD$'=$JD-2450000.
}
\end{figure}

	If we assume no blending (see \S\ 5), then $\overline A(\bu_c)$ is directly
measured from the light curve (Fig.\ \ref{fig:one}).  In principle,
$A_3$ is changing with time and so can only be known if there is a full
model of the light curve.  However, in general $A_3$ is expected to
change slowly.  Moreover, in the present case it is measured 
beginning at JD$'=1733.65$,
only 2.5 hours after the last VLT measurement (JD$'=1733.54$) and is found
to be nearly constant for an hour after that.  We will therefore assume that
$A_3$ is constant throughout the crossing and will discuss possible small
corrections to this approximation in \S\ 5.

	For the last VLT point, Albrow et al.\ (2001b) report
$\overline{E}(1733.54) = 0.79\pm 0.03\,$\AA, and they adopt $E_0=1.03\,$\AA,
i.e. $\Delta =-0.23 \pm 0.03$.  From the EROS lightcurve (Fig.\ \ref{fig:one}),
$f(1733.54) = 0.060\pm 0.007$.  As we will show in \S\ 4, at this time
the source center was approximately $\eta = 0.963$ source radii past the
caustic, so that $\overline{E_2}$(1733.54) is a weighted average of the outer
4\% of the source.  Here $\eta$ is the perpendicular distance from the
source center to the caustic in units of the source radius, and is taken
to be negative when the source center is inside the caustic.
Combining these facts with equation (\ref{eqn:h2eval})
yields
\begin{equation}
{\overline{E_2}(1733.54)\over E_0} = -2.9\pm 0.7\qquad (\eta\simeq 0.96),
\label{eqn:h2evalN5}
\end{equation}
which is to say that the limb is in emission more strongly than the
star as a whole is in absorption.  As we show in \S\ 4, 
this would strongly contradict the model of Albrow et al.\ (2001b).

	We note that since the caustic is a square-root singularity
(and assuming circular symmetry of the source),
\begin{equation}
\overline{E_2} \simeq 
{\int_\eta^1 dr rS(r)E(r)\over \int_\eta^1 dr rS(r)}\qquad (1-\eta\ll 1),
\label{eqn:h2evalapprox}
\end{equation}
where $r$ is normalized to the source radius.
That is, $\overline{E_2}$ is very nearly equivalent to a simple weighted
average of the EW
by the light coming from the {\it entire} periphery of the source.

	Applying equation (\ref{eqn:h2eval}) to all the VLT and Keck
measurements, we find,

\vspace{1cm}
\tablecaption{Implications of Measured H$\alpha$ EWs}
\tablehead{\hline\hline}
\tabletail{\hline}
\tablefirsthead{
{Telescope} & 
{JD$'$} & 
{$\overline{E}$ (\AA)} & 
{$\sigma(\overline{E})$ (\AA)} & 
{$f$} & 
{$\sigma(f)$} & 
{$\eta$} & 
{$\overline{E_2}$ (\AA)} & 
{$\sigma(\overline{E_2})$ (\AA)} \\ 
\hline}
\begin{supertabular}{l l l l l l l l l}
VLT &  1730.60  & 1.06  & 0.01  & 0.846 & $0.001$ & $ -0.595$ & $ + 1.07$ & 0.01 \\
VLT &  1731.67  & 1.09  & 0.02  & 0.803 & $0.001$ & $ +0.097$ & $ + 1.10$ & 0.02 \\
VLT &  1732.66  & 0.98  & 0.01  & 0.626 & $0.001$ & $ +0.611$ & $ + 0.95$ & 0.02 \\
VLT &  1733.54  & 0.79  & 0.03  & 0.060 & $0.007$ & $ +0.963$ & $  -2.97$ & 0.68 \\
Keck & 1731.953 & 0.944 & 0.002 & 0.773 & $0.001$ & $ +0.256$ && \\              
Keck & 1732.950 & 0.869 & 0.006 & 0.510 & $0.002$ & $ +0.738$ && \\
\end{supertabular}

\vspace{1cm}

Note that since Castro et al.\ (2001) did not measure $E_0$,
we cannot use equation (\ref{eqn:h2eval}) to evaluate $\overline{E_2}$,
for the Keck data.  One cannot simply adopt the VLT value of $E_0$
because the VLT and Keck data are affected by systematic differences 
as discussed in the penultimate paragraph of \S\ 4.
We note that D.\ Minniti (2001, private
communication) has high-resolution VLT data for this event when it
was highly (but not differentially) magnified, so it may eventually be
possible to correct this shortcoming.

	The results reported in Table 1 can be directly compared with results
from model atmospheres, subject to some minor qualifications discussed
in \S\ 5.  

\section{Predicting Equivalent Widths from the Lightcurve}	

	Without loss of generality, one can write the magnification as
\begin{equation}
A(\bu) = A_3(\bu) + Z(\bu)\Delta u_\perp^{-1/2} \Theta(\Delta u_\perp),
\label{eqn:ageneral}
\end{equation}
where $\Delta u_\perp$ is the perpendicular distance from $\bu$ to the
caustic and $\Theta$ is a step function.  As we discussed earlier, unless
the source is actually probing the cusp region, it is an excellent 
approximation to substitute $A_3(\bu)\rightarrow A_3(\bu_c)$.  Since
the cusp approach occurred $\sim 4\,$days after the caustic exit, and
since $A_3(\bu_c)$ was observed to be almost perfectly flat for an hour after
the crossing, we adopt this approximation.  For closely related reasons
(discussed in \S\ 5), we
adopt $Z(\bu)\rightarrow Z(\bu_c)$.  Under these assumptions, and
making use of the fact (explicitly demonstrated in \S\ 5) that the 
caustic curvature is small on scales of the source, the spatial averaging over
the model atmosphere (required to compare theoretical predictions with 
the empirical results summarized in Table 1) is independent of $Z$,
\begin{equation}
\overline{A_2}(\eta)\rightarrow Z(\bu_c) G(\eta),\qquad
G(\eta) = {1\over F_s}\int_{\eta}^1 dx (x-\eta)^{-1/2}
\int_{-\sqrt{1-x^2}}^{\sqrt{1-x^2}}dy S(r),
\label{eqn:geval}
\end{equation}
\begin{equation}
\overline{E_2}(\eta)\rightarrow {H(\eta)\over G(\eta)},\qquad
H(\eta) = {1\over F_s}\int_{\eta}^1 dx (x-\eta)^{-1/2}
\int_{-\sqrt{1-x^2}}^{\sqrt{1-x^2}},
dy E(r) S(r)
\label{eqn:geval2}
\end{equation}
where $x$ and $y$ are dimensionless variables and
$r\equiv\sqrt{x^2+y^2}$.  Note that for a uniform source, $G(\eta)$
is given by e.g., Figure 1 of Gould \& Andronov (1999).

	In this section we will make the additional assumption that
$A_3$ and $Z$ are independent of $\bu_c$,
\begin{equation}
A_3(\bu_c) \rightarrow A_3,\qquad Z(\bu_c)\rightarrow Z\qquad (\rm assumption).
\label{eqn:a3zconst}
\end{equation}
As we discuss in \S\ 5, this assumption is less robust than others made
in this paper, but is still quite reasonable for our purposes.

	Albrow et al.\ (2001b) have presented calculations of $A(\eta)$ and 
$E(\eta)$ for a K3 giant atmosphere model in their Figures 1 and 4
respectively, under the assumption that equation (\ref{eqn:a3zconst}) is
valid, for two specific choices of the pair $(A_3,Z)$.  The resulting curves 
were intended only for schematic purposes, and in particular assumed that
$\eta$ was a linear function of time [$\eta=({\rm JD}' - 1731.8)/1.9$],
which is known not to hold in the present case (see eq.\ [\ref{eqn:parab}],
below).  However, our interest in these curves is solely that they allow
us to evaluate the functions $G$ and $H$ that derive from the underlying
atmosphere model.  As a matter of practical
computation, we must normalize $G$, and so we adopt $G_{\rm max}=1.40.$
However, the final predictions of EW are completely invariant under
changes of this parameter.

\begin{figure}
\epsfxsize \hsize
\epsfbox{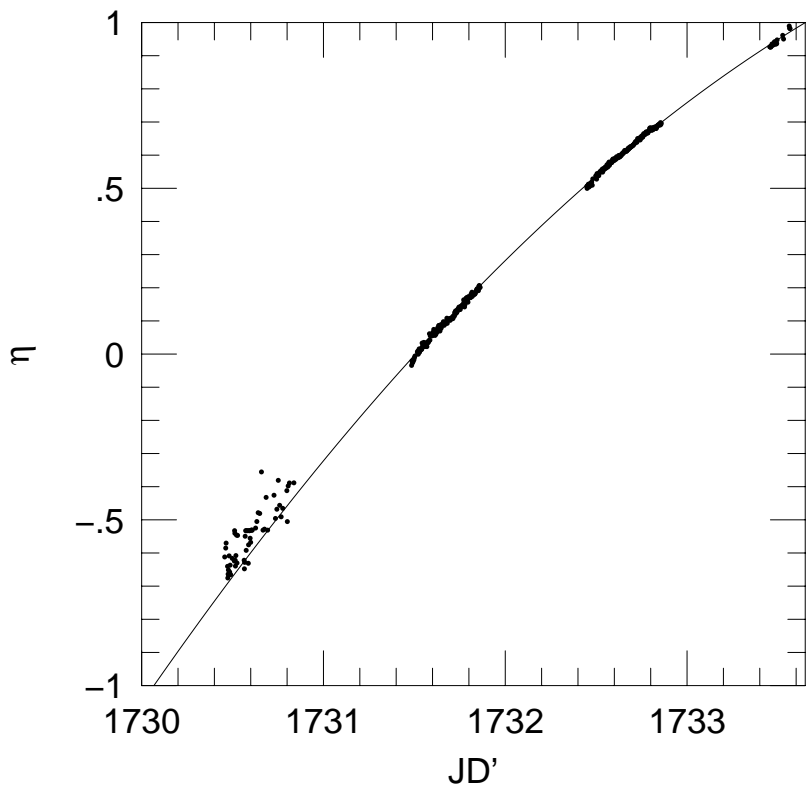}
\caption[junk]{\label{fig:two}
Photometric determination of the perpendicular distance $\eta$ from the 
source center to the caustic in units of the source radius.  First, we find
$G(\eta)=(A-A_3)/Z$ where $A$ is the observed magnification 
(Fig.\ \ref{fig:one}), $A_3=3.23$ is the post-caustic magnification,
and $Z=(A_{\rm max} -A_3)/G_{\rm max}=12.7$ (see text).  We then
invert the stellar-profile function $G(\eta)$ from Fig.\ 1 of Albrow et al.\ 
(2001b) to determine $\eta$.  The curve is a quadratic fit to all the points.
The fit time of its endpoint (JD$'=1733.65$ at $\eta=1$) occurs just before
the start of the last hour of data which, from Fig.\ \ref{fig:one}, is
independently known to occur after the caustic exit.
}
\end{figure}

	We then apply these extracted functions to the EROS data.  First,
we measure $A_3 = 3.2294\pm 0.0034$, where the error includes only the
measurement error of the post-caustic flux (and not the baseline flux
that normalizes it).  This is because all predictions are invariant under
changes of the assumed baseline flux.  We also measure
$\overline{A}_{\rm max}=21.02$ from the peak of the curve,
and hence
find $Z = (\overline{A}_{\rm max}-A_3)/G_{\rm max} =12.7$.  Then for each
observed magnification, $\overline{A}$, we determine 
$\eta=G^{-1}[(\overline{A}-A_3)/Z]$.  Since $G^{-1}$ is
double valued we must assign a branch depending on whether JD$'$ is before
or after 1730.59 (the time of the observed peak of the light curve), 
and for points that have 
$\overline{A}>\overline{A}_{\rm max}$, due to observational error, we
assign $\eta=G^{-1}[(\overline{A}_{\rm max}-A_3)/Z]$.  The resulting values
of $\eta$ are shown in Figure \ref{fig:two}.  For most caustic crossings,
$\eta(t)$ is reasonably approximated by a straight line, but clearly
this is not so in the present case.  Indeed from the cusp approach 
(sharp peak) in the light curve four days after the caustic crossing, it is
known that the source must have exited nearly tangent to the caustic, so that
the caustic curvature must play an important role.
To interpolate to times
when we have no data, we fit therefore the points 
$\eta(t)$ to a parabola whose equation
is given by,
\begin{equation}
\eta = 1 + 0.3298({\rm JD}' -t_{\rm cc,end} ) 
- 0.0637({\rm JD}' -t_{\rm cc,end} )^2,
\qquad t_{\rm cc,end}=1733.6506,
\label{eqn:parab}
\end{equation}
and whose form is shown in Figure \ref{fig:two}.  This curve is a
good fit to the data except for the first night where it deviates by
approximately the scatter in the points.  This shows that at least one
of our assumptions is not fully satisfied: either the caustic deviates
from a parabolic form, or the assumption (\ref{eqn:a3zconst}) is too
much of an oversimplification, or perhaps the limb-darkening model
of Albrow et al.\ (2001b) does not perfectly reproduce $G(\eta)$ for the
EROS $V_E$ band.  Whatever the {\it cause} of the deviation, 
the fact that it is more pronounced on the first night is simply due to the
fact that the errors in $\eta$ are larger.  This in
turn can be traced to the fact that this night covers the peak of the
curve (see Fig.\ \ref{fig:one}), where $dG/d\eta\sim 0$.  The error
in the estimate of $\eta$ for each measured point is given by
$\sigma(\eta)=\sigma(F_V)/(F_{V,\rm baseline}Z|dG/d\eta|)$.  These
larger errors are reflected in the larger scatter of the points
on the first night in Figure \ref{fig:two}.  The errors are not displayed
in Figure \ref{fig:two} in order
to avoid clutter, but are shown in a different projection in \S\ 5.

	Note that the error on the time of the
end of the caustic crossing ($\eta=1$), derived from the fit
leading to equation (\ref{eqn:parab}), is very small,
$t_{\rm cc,end}= 1733.6506\pm 0.0048$.  However, the actual error is dominated
by the assumption that the limb-darkening model of the curve shown
in Figure \ref{fig:one} (ultimately derived from Fig.\ 1 of Albrow et al.\
2001b)
is correct, which it almost certainly is not
at this level.  Nevertheless, $t_{\rm cc,end}$ cannot be much after this
best-fit value because the observed light curve is flat 
(or very slightly rising) beginning at this time.

	Finally, we predict the EW, $\overline{E}(t)$, using the 
values of $\eta(t)$
found from equation (\ref{eqn:parab}), the tabulated functions
$G(\eta)$ and $H(\eta)$, the measured values of $\overline{A}(t)$ 
and $A_3$, and equations (\ref{eqn:hbar}) and (\ref{eqn:geval2}).
The result is shown in Figure \ref{fig:three}.  

	As can be seen
from Table 1, the VLT EWs are about 0.1 \AA\ larger than the 
corresponding Keck EWs, even taking account of the fact that the
observations were a few hours earlier.  One possible explanation for
this discrepancy is that the low-resolution VLT spectra were affected
by blended lines that were removed from the Keck HIRES spectra.  From
Figure 2 of Castro et al.\ (2000), this indeed appears plausible.
For purposes of displaying our results in Figure \ref{fig:three}, we therefore
assume that 0.1 \AA\ of the VLT EWs are due to blends.  Hence, we 
reduce all of them by this amount and correspondingly rescale the
Albrow et al.\ (2001b) model downward by 10\%.  There remains the
freedom to adjust the relative offsets between the Keck and VLT
points (because our 10\% VLT blending estimate is only a guess), and
to adjust the overall scale of the curve (because this is set by $E_0$
which is measured from the VLT pre-caustic spectrum -- but with a large
error bar).

	Nevertheless, there is no amount of adjustment that
can reconcile the points with the curve.
The point with the largest deviation between observed and predicted
EW is the last VLT spectrum.  In the
previous section we showed that the model could predict such
a low EW only if the outer 4\% of the source were strongly in 
emission.  By contrast, the model of Albrow et al.\ (2001b)
predicts that the outer 4\% will have substantial (albeit reduced) 
{\it absorption}:
$\overline{E_2}(0.963) = H(0.963)/G(0.963)=0.0105\,$\AA/0.0184=0.57\,\AA\ 
(with no correction for line blends).

\section{Examination of Assumptions}

	We have used EROS lightcurve data to facilitate the comparison
of spectra taken during the caustic crossing of EROS-BLG-2000-5 with
an atmosphere model of the K3 source.  An alternative approach would
be to fit the lightcurve data to a lens model and then apply the
lens model directly to the atmosphere model to make predictions about
the spectra.  The latter approach would appear to be more secure but,
unfortunately, to date there are no published models of this event, and
the event may prove difficult to model.  Thus, since our approach is
the only one available at present, it will be worthwhile to go through
the assumptions that were introduced and see what order of errors they
induce.  This review can also serve as a checklist if our method is
applied to future events.

	We assumed no blending.  There are three lines of evidence
that this is a good assumption.  First, the lensed source is a bright 
giant, so it is fairly improbable that it would be seriously blended.
Second, there is no evidence for blended light in the spectrum of
Castro et al.\ (2001) as there would be if there were a bright blend
offset by at least $15\,\kms$ in radial velocity from the source.
Third, the event is nearly achromatic (see Fig.\ \ref{fig:four}).
This means that either the blend is very weak
in which case it has little effect, or it has virtually the same 
color as the source, and so very likely has the same H$\alpha$ EW as well.
To the extent that this is the case, the predicted EWs are exactly the
same as in the unblended case.

	Next, we assumed that the non-divergent magnification field 
$A_3(\bu)$ could be replaced by its value at the center of the source,
$A_3(\bu_c)$.  This is valid to the extent that on scales of the source
size, the $A_3$ field is equal to a constant plus a gradient.  The only way
this could fail to hold would be if the source were probing the
neighborhood of a cusp.  It is clear that the source {\it is}
approaching a cusp, because the lightcurve betrays a characteristic 
cusp-approach behavior 4 days after the caustic exit.  However, in the
hour after the caustic exit, the light curve is nearly flat,
indicating that this cusp is {\it not yet} significanly affecting
the magnification pattern at the time of the exit.

	Third, we assumed that $A_3$ would be constant throughout
the crossing.  The measured logarithmic slope of $A_3$ immediately
after the crossing is $d\ln A_3/d t= (0.11\pm 0.08)\,\rm day^{-1}$, so that 
at the time of the last spectrum measurement (0.11 days before the end of the 
crossing),
$A_3$ was within a percent or two of the final value.
A 1\% change in $A_3$ leads to a change of $\sim -17\%$ in $f$ and so
of $\sim -0.5$ in equation (\ref{eqn:h2evalN5}).  This is not negligible,
but it would not qualitatively affect the conclusion that the outer
4\% of the source is in emission.
Plausibly, at much earlier times $A_3$ could have differed 
from its final value by $\sim 10\%$, but as these times such a change 
would have a very small impact on the conclusions simply because
$A_3$ does not dominate the total flux.  For example, suppose this
had been the case at JD$'=1732.66$.  Then $f$ would have fallen
from 0.626 to 0.588, which in turn would change $\overline{E_2}$ by
only 0.005 \AA.  These are all the assumptions needed through
\S\ 3.

	In \S\ 4, we went on to make three additional assumptions.
The first was that on the scale of the source
the caustic could be approximated as a
straight line.  From the measurements of the first crossing, the
source is known to cross its own diameter in $\la 1\,$day.
If we interpret the quadratic behavior of equation
(\ref{eqn:parab}) as arising from the curvature of the caustic,
then over a diameter, the caustic deviates from a straight line
by $\sim 8\times 10^{-3}$ of a source radius, certainly small
enough to ignore.

	Second, we assumed that at any instant $Z(\bu)$ could
be replaced by its value at its center $Z(\bu_c)$.  Or equivalently, that
we are in the square-root singularity regime and the
coefficient of this regime varies at most linearly over a source
diameter.  From Table 1 and the adopted value of $Z=12.7$, one sees that
that at the trailing limb of the source for the first measured point,
$A_2(\bu) = 10$, which is normally well into the square-root singularity.
Since the source diameter is $\la 1\,$day and at the exit it is still
4 days from the post-caustic peak in the light curve due to a cusp approach, 
the assumption that quadratic variation
of $Z$ is not significant on the scale of the source size appears very
reasonable, but by no means proven.

\begin{figure}
\epsfysize 15cm
\epsfbox{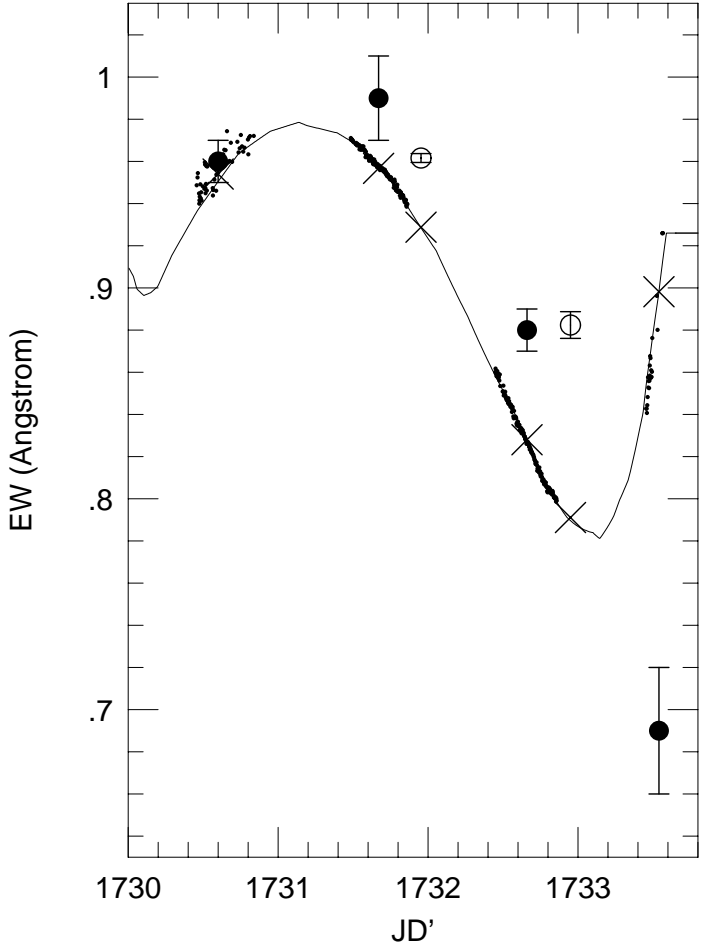}
\caption[junk]{\label{fig:three}
Measured and predicted H$\alpha$ equivalent widths (EWs).  Small points show
the predictions from the photometric data.  For each measured magnification
(Fig.\ \ref{fig:one}), the source-caustic separation $\eta$ is determined
(Fig.\ \ref{fig:two}), and so also the EW of the two highly magnified images
inside the caustic, $\overline{E_2}=H(\eta)/G(\eta)$.  Here $G$ and $H$ are 
functions extracted from the atmosphere model of Albrow et al.\ (2001b).  
See text.  The predicted EW is then the weighted average of $\overline{E_2}$
and $E_0$, the EW of the unmagnified source.  The curve results from applying
the same calculation to the curves in Figs.\ \ref{fig:one} and \ref{fig:two}.
Solid points are measured EWs from VLT FORS1 low-resolution data of 
Albrow et al.\ (2001b) and open points are from HIRES KECK data of 
Castro et al.\ (2001).  Crosses show where these points should be if they
agreed with the model.  The VLT points and the corresponding model have
been rescaled assuming a 10\% blend from a neighboring line.  This
is plausible based on Fig.\ 2 from Castro et al.\ (2001) and brings the
VLT points approximately into line with the Keck points.
However, no amount of rescaling would bring the atmospheric model into
agreement with all the data points.
}
\end{figure}

\begin{figure}
\epsfxsize \hsize
\epsfbox{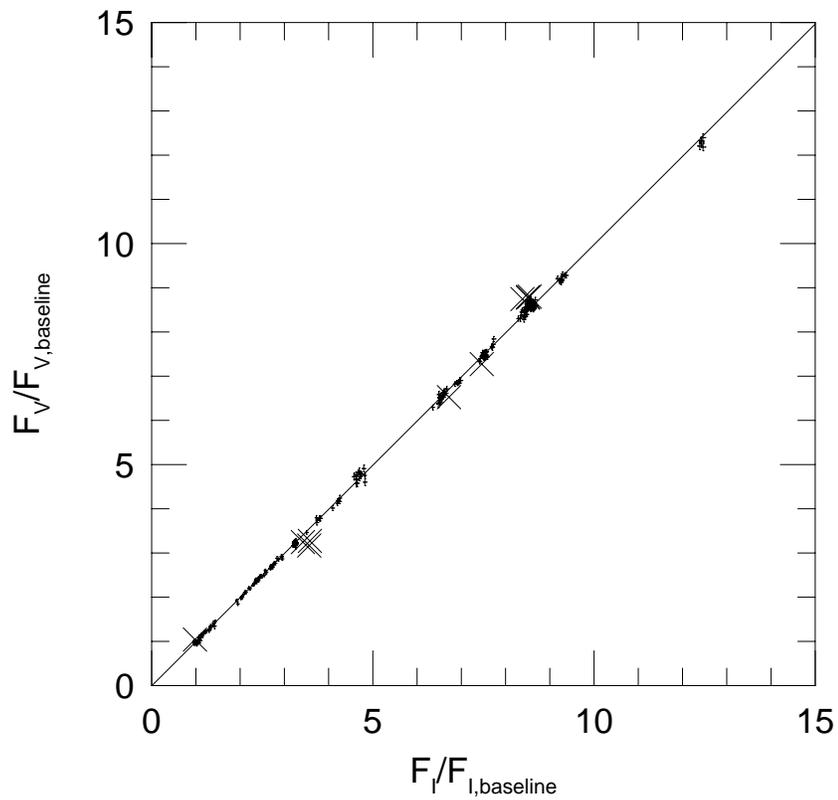}
\caption[junk]{\label{fig:four}
Comparison of fluxes in two bands $V_E$ and $I_E$ for all points on the
EROS light curve except during the two caustic crossings and the
cusp approach (when differential magnification could induce color changes).
The regression (solid line) passes extremely close to the origin, showing
that either the event is not significantly blended or if it is, the
color of the blend is extremely close to that of the source.
The large crosses show the 10 outlier points (out of 250) that were excluded
from fit.
}
\end{figure}

	Third, we assumed that $A_3$ and $Z$ remain constant throughout
the crossing.  We discussed above why this is a reasonable assumption
for $A_3$.  The argument regarding $Z$ is similar.  Suppose that
$Z$ in fact fell linearly by 20\% from the time of the peak until
the end of the crossing (and suppose $A_3$ remained constant).
The product $G(\eta)Z=\overline{A_2}$ is then an observable and so
remains fixed.  Thus, at the last measurement, $Z$ would be 19\% lower,
so $G(\eta)$ would actually be 24\% higher than we have allowed.  This
would drive $\eta$ down from 0.963 to 0.956, but it would affect the
EW prediction shown in Figure \ref{fig:three} by well under 1\%.  
Hence, while it would be better not to have to make this approximation, 
the errors it induces are extremely small compared to the level of 
disagreement between the models and the data in Figure \ref{fig:three}.

\begin{figure}
\epsfxsize \hsize
\epsfbox{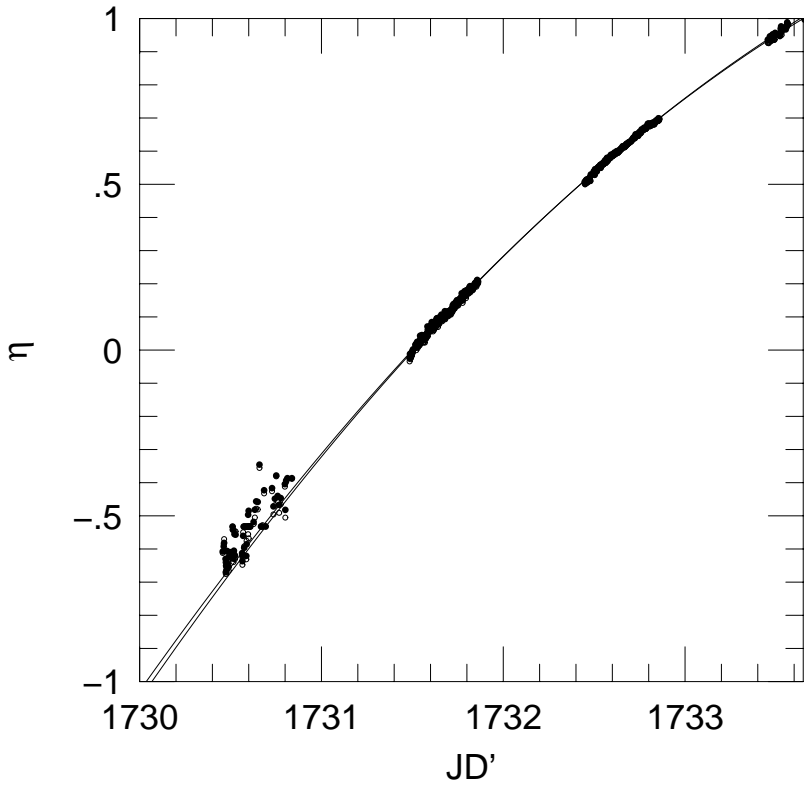}
\caption[junk]{\label{fig:six}
Filled points and upper curve:
Photometric determination of the perpendicular distance $\eta$ from the 
source center to the caustic in units of the source radius, obtained
by directly fitting the EROS $V_E$ data to a limb-darkened profile function
$G_V(\eta;c)$,
where the linear limb-darkening parameter $c$ is treated as a free parameter
and is found to be $c=0.644\pm0.032$.
Open points and lower curve: Same as Fig.\ \ref{fig:two}, 
which was based on the stellar
profile function $G_P(\eta)$ taken from Albrow et al.\ (2001b).
The a priori model $G_P(\eta)$ yields almost exactly the same curve
as the limb-darkening fit $G_V(\eta;c)$.
}
\end{figure}

	To directly probe the scale of the systematic errors
introduced by our approximations, we conduct the following test.
Instead of adopting the $G_P(\eta)$ derived from the atmosphere model
of Albrow et al.\ (2001b), we directly fit for $G_V(\eta;c)$
from the observed $V_E$ magnification, $A_V(t)$, using the 6-parameter
functional form,
\begin{equation}
A_V(t) = A_3 + ZG_V[\eta(t);c],
\label{eqn:limbdark}
\end{equation}
where $c$ is a linear limb-darkening coefficient, and $\eta$ is
a (3-parameter) quadratic function of time as in equation
(\ref{eqn:parab}).  We find $c=0.644\pm 0.032$, $t_{\rm cc,end}=1733.6280$,
and linear and quadratic coefficients $0.3421$ and $-0.0596$, respectively.
The resulting curve $\eta(t)$ (Fig.\ \ref{fig:six})
is very similar to the corresponding curve (Fig.\ \ref{fig:two}) for $G_P$.
Moreover, when we evaluate $\eta$ using $G_V$ instead of $G_P$, we
find that the predicted EWs are almost exactly the same as those shown
in Figure \ref{fig:three}: the changes are an order of magnitude smaller
than the error bars on the VLT and Keck data points.

\begin{figure}
\epsfxsize \hsize
\epsfbox{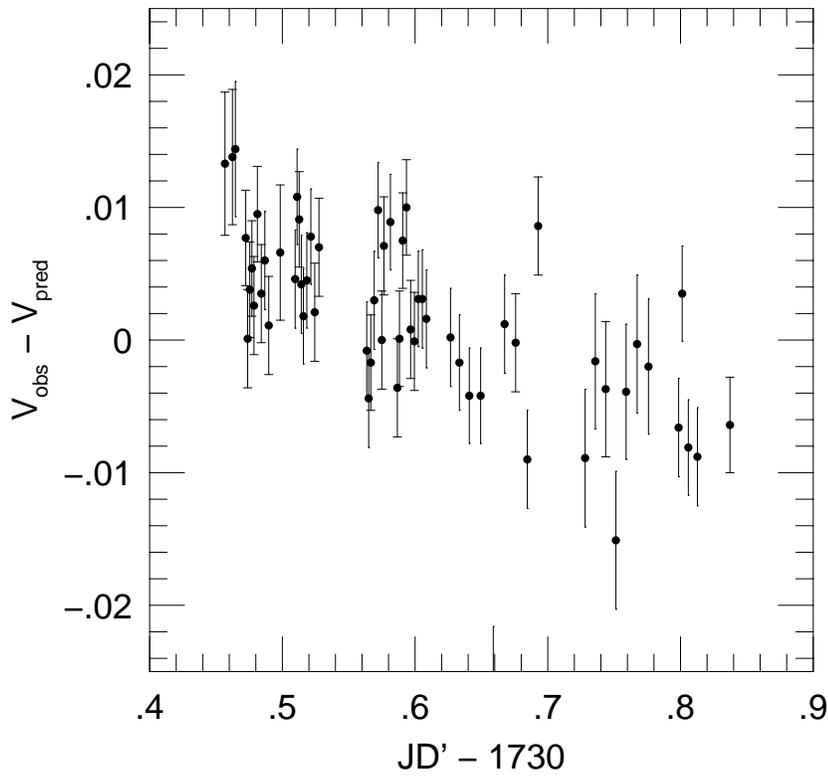}
\caption[junk]{\label{fig:seven}
Photometric ($V_E$) residuals on the first night of the caustic crossing,
relative to the linear limb-darkening fit, with a parabolic $\eta(t)$
shown in Fig.\ \ref{fig:six}.  The individual residuals are extremely
small, but the trend is highly significant, showing that the real light
curve peaks earlier than the model by about 0.1 days.
}
\end{figure}

	Figure \ref{fig:seven} shows the {\it photometric} residuals 
of the fit during the first night (JD$'\sim 1730.6$), for which the residuals
in terms of $\eta$ seen in Figure \ref{fig:six} are most pronounced.
Typically, these residuals are extremely small, $<0.005$ mag, and 
are of order the individual errors.  However, taken
together they show a clear trend which is highly significant: the observed
light curve peaks earlier than the model.  The offset is about 0.1 day,
or $\Delta\eta\sim 1/15$.  This level of discrepancy induces an offset
in the predicted EW on the first night of about 1/2 the statistical error
of the EW measurement (see Fig.\ \ref{fig:three}).  However, since the
model is only used to interpolate to times when there are no photometric
data, and since the model and data are in excellent agreement in
the neighborhoods of the Keck points (which are the only ones where
there is no coincident photometry), this discrepancy has no practical
impact.

	Parenthetically, we note that while one might be tempted to
take the good agreement shown in Figure \ref{fig:six} as supporting
the believability of the derived limb-darkening coefficient, in
fact this determination must be viewed skeptically.  Any change
in $A_3$ or $Z$ during the very long crossing, as well as any deviation
of the caustic from a purely parabolic form, will be subsumed in
the limb-darkening coefficient.  Hence, a valid measurement of $c$
can be obtained only in conjunction with a complete determination
of the geometry of the binary lens.  The main implication of this
fit is that, while the limb-darkening coefficient may be significantly
affected by these various deviations from a simple model, the
predictions of equivalent widths are not.

	We conclude that all the assumptions we have made are quite
reasonable, at least for the present case.  Future applications of
this method will have to be carefully justified based on an analysis
of the event that is parallel to the one given here.

\section{Color of the Limb}

	Essentially the same method presented above can be used to
find the color of the two highly magnified (caustic) images.  Recall
from equation (\ref{eqn:h2evalapprox}) that for $1-\eta\ll 1$, this is 
very nearly equivalent to the color of the annulus of the source
$1-\eta < r < 1$.  The difference
in the color of the two highly magnified images relative to the unmagnified
source is given by,
\begin{equation}
\Delta (V_E-I_E)\equiv (V_E-I_E)_2 - (V_E-I_E)_{\rm source} = 
2.5\,\log\biggl({F_{I,2}\over F_{V,2}}\,{d F_V\over d F_I}
\biggr)
\label{eqn:deltacol}
\end{equation}
where $F_{I,2}$ and $F_{V,2}$ are the fluxes due to the two caustic images, 
and $d F_V/ d F_I$ is
the slope shown (normalized to the baselines) in Figure \ref{fig:four}.  
Since this figure is restricted
to points away from the caustics and cusps where the source is not 
differentially magnified, $-2.5\log(d F_V/ d F_I)$ is the
color of the source.  Under the assumption (\ref{eqn:a3zconst}),
\begin{equation}
F_{V,2} = F_V - F_{V,\rm end},
\label{eqn:a3zconst2}
\end{equation}
where $F_V$ is the observed $V_E$ flux at a given time and $F_{V,\rm end}$
is the flux just after the end of the caustic crossing (and
similarly for $I_E$).  Note
that equations (\ref{eqn:deltacol}) and (\ref{eqn:a3zconst2}) are valid 
independent of whether the source is blended, and that when combined
they can be evaluated from directly observed quantities without any
modeling.  The only assumption is that $A_3$ does not change significantly
between the time of the measurement and the end of the caustic crossing.

\begin{figure}
\epsfxsize \hsize
\epsfbox{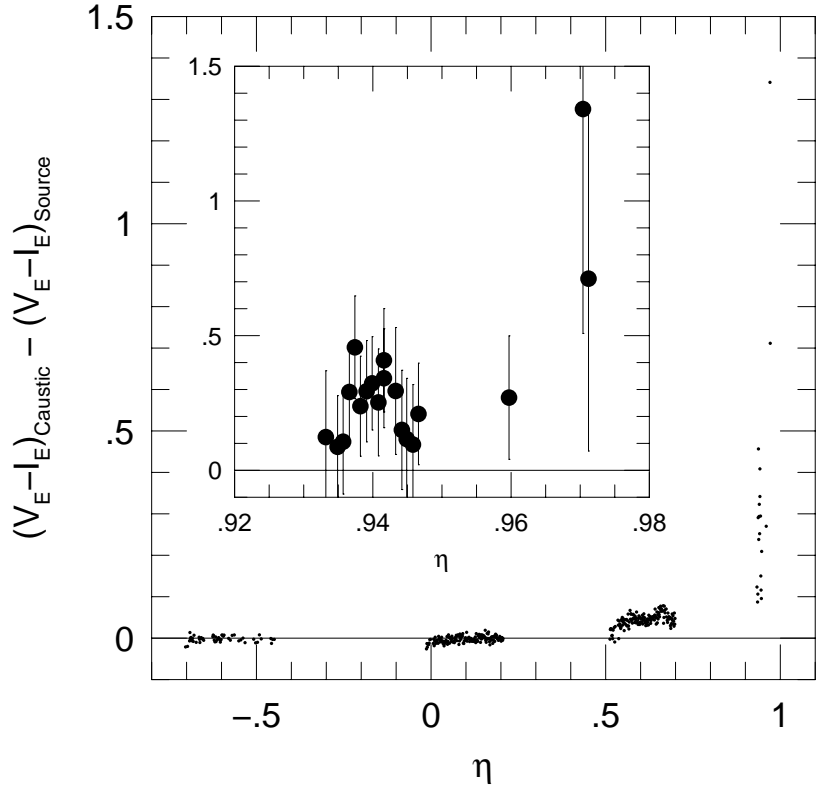}
\caption[junk]{\label{fig:five}
Difference in $(V_E-I_E)$ color between the two highly-magnified caustic
images and the source as a whole, as a function of $\eta$, the separation
of the source center from the caustic in units of the source radius.
The color shift is extremely small and only marginally significant on
the first two nights, but rises to $0.046\pm 0.001$ on the third night
and $0.27\pm 0.02$ on the last, when
the caustic images are equivalent to looking at
the outer 5.5\% of the source.  Since the source is a K3 giant, this
limb has the color of a M0 giant.
}
\end{figure}

	Figure \ref{fig:five} shows the result of applying equations
(\ref{eqn:deltacol}) and (\ref{eqn:a3zconst2}) to the EROS data.
The time coordinate has been changed to source position relative
to the caustic $(\eta)$ using equation (\ref{eqn:parab}).
The color difference is small to negligible on the first two nights
$[\Delta(V_E-I_E)=0.0006\pm 0.0006$ and $-0.0015\pm 0.0006$ for
$\eta\sim -0.6$ and $\eta\sim 0.1$ respectively] reflecting the fact
that the caustic images receive significant contributions from a range
of radii at these positions (Gaudi \& Gould 1999).  On the third night,
the caustic images are significantly redder, 
 $\Delta(V_E-I_E)=0.046\pm 0.001$ at $\eta\sim 0.6$.  Here, the
analog of equation (\ref{eqn:h2evalapprox}) approximately holds, so
that the color of the caustic images is the same as that of the
outer $1-\eta\sim 40\%$ of the source (by radius).  Finally, on the last night,
we find $\Delta(V_E-I_E)=0.27\pm 0.02$ at $\eta\sim 0.945$, equivalent
to an annulus comprising the outer 5.5\% of the source.  This is
roughly equivalent to $\Delta(V-I)\sim 0.37$ in standard Johnson/Cousins
bands.  Hence, assuming that the K3 source has a typical $(V-I)_0\sim 1.36$
color (Bessell \& Brett 1988), the limb has $(V-I)_0\sim 1.73$, i.e. the
color of a M0 giant.


\section{Probing the Chromosphere?}

	The color (and thus temperature) difference of the source periphery
relative to the source as a whole ($\Delta(V-I)\sim 0.37$) is 
modest compared to the dramatic shift from H$\alpha$ absorption to emission
implied by equation (\ref{eqn:h2evalN5}).  That is, M0 stars absorb rather
than emit H$\alpha$.  Hence, to explain this shift it would appear necessary
to invoke some additional {\it structure} in the source atmosphere.
The chromosphere presents itself as an interesting candidate.  We therefore
evaluate what strength of chromospheric emission is required to account
for the sharp decline in EW reported by Albrow et al.\ (2001b) on the
last night.

	Seen in projection
above the limb of the photosphere, the solar chromosphere emits strongly
in H$\alpha$ and is essentially black in the neighboring continuum.  If
we assume that the same is true of the source of EROS-BLG-2000-5, then
equation (\ref{eqn:hbar}) can be generalized to
\begin{equation}
\overline{E}(\bu_c) 
= {A_3(\bu_c)E_0 + \overline{A_2}(\bu_c)\overline{E_2}(\bu_c)
-\overline{A_{2,\rm chrom}}(\bu_c)E_0 X
\over A_3(\bu_c) + \overline{A_2}(\bu_c)},
\label{eqn:hbar3}
\end{equation} 
where $E_0$ still represents the EW of the unmagnified source, including 
the effect of H$_\alpha$ chromospheric emission, X is the ratio of this 
chromospheric emission to the algebraic sum of the H$_\alpha$ absorption 
by the stellar atmosphere and of the chromospheric emission,
and 
$\overline{A_{2,\rm chrom}}(\bu_c)$ is the magnification of the chromosphere
by the two caustic images.  
Note that $\overline{A_{2,\rm chrom}}$ is
the  ratio of the H$_{\alpha}$ chromosphere photon flux 
in the two caustic images
to the H$_{\alpha}$ chromosphere photon flux 
normally received in the absence of magnification, 
 and so is not exactly analogous to
 $\overline{A_{2}}$, which is normalized to the number of
photons normally received from the source {\it as a whole}.

	It is straightforward to show that in the limit $\eta\rightarrow 1$,
\begin{equation}
\overline{A_{2,\rm chrom}}\rightarrow {Z_*\over \sqrt{2}}
\qquad (\eta\rightarrow 1).
\label{eqn:achrom}
\end{equation}
See for example Figure 1 from Castro et al.\ (2001).  Here, $Z_*$ is the
magnification due to the two caustic images of a point source lying exactly
one source radius inside the caustic.  We will adopt $Z_*=12.7$, which
is the value measured in \S\ 4.  As discussed in \S\ 5, $Z$ is
actually a function of $\bu_c$ and so in principle could be different at the
peak of the light curve $(\eta\sim -0.5)$ where it is measured, from its
value at $\eta=0.963$.  We therefore assign this parameter an error of 20\%.
(In addition, an error could be introduced if $G_{\rm max}$ for the actual
limb-darkened profile differed from the one adopted.  However, for a 
reasonable range of limb-darkening profiles, this uncertainty is $<2\%$.)\ \ 
Solving equation (\ref{eqn:hbar3}) for $X$ and substituting in values
for the various parameters, we find
\begin{equation}
X = \sqrt{2}{A_3\over Z_*}
\biggl[
\biggl(1- {\overline{E}\over E_0}\biggr) 
-{\overline{A_2}\over A_3}\biggl(
{\overline{E}\over E_0} - {\overline{E_2}\over E_0} 
\biggr)\biggr] = 0.08 \pm 0.02,
\label{eqn:xeval}
\end{equation}
where we have taken $\overline{E_2}/E_0=0.57$ as evaluated in \S\ 4 
based on the
Albrow et al.\ (2001b) model.  Note that since the ratio of the two
terms within the brackets is $\sim 0.05$, uncertainties in the estimate of
$\overline{E_2}/E_0$ are not
likely to significantly impact the final result.

	This value of $X$ is about two orders of magnitude higher than
the corresponding quantity for the Sun, but giant stars may have a 
more significant chromosphere than the Sun.  Alternatively, the
errors in the H$\alpha$ measurement may have been underestimated.
In any event, this estimate of $X$ can be directly compared with chromosphere 
models.

\begin{acknowledgements}:
We are grateful to D.\ Lacroix and the technical staff at the Observatoire de
Haute Provence and to A.\ Baranne for their help with the MARLY
telescope.  We are also grateful for the support
given to our project by the technical staff at ESO, La Silla. We thank
J.F.\ Lecointe and A.\ Gomes for assistance with the online computing.
We thank G.\ Newsom for useful discussions on the solar chromosphere.
Work by AG was supported by NSF grant AST~97-27520 and by
a grant from Le Centre Fran\c cais pour L'Accueil et Les \'Echanges
Internationaux.
\end{acknowledgements}

\clearpage

\clearpage 

\begin{thebibliography}{} 

\bibitem{} Afonso, C.\ et al.\ 2000, ApJ, 532, 340
\bibitem{} Alard, C.\ 2000, A\&AS, 144, 363
\bibitem{} Albrow, M.\ et al.\ 1999, ApJ, 522, 1011 
\bibitem{} Albrow, M.\ et al.\ 2000, ApJ, 534, 894 
\bibitem{} Albrow, M.\ et al.\ 2001a, ApJ, 549, 759
\bibitem{} Albrow, M.\ et al.\ 2001b, ApJ, 550, L173
\bibitem{} Alcock, C., et al.\ 1997, ApJ, 491, 436 
\bibitem{} Ansari, R., et al.\ 1996, Vistas in Astronomy, 40, 519
\bibitem{} Bessell, M.S., \& Brett, J.M.\ 1988, PASP, 100, 1134
\bibitem{} Castro, S.M., Pogge, R.W, Rich, R.M., DePoy, D.L., 
\& Gould, A.\ 2001, ApJ, 548, L197
\bibitem{} Gaudi, B.S., \& Gould, A.\ 1999,,ApJ, 513, 619
\bibitem{} Gould, A., \& Andronov, N.\ 1999, ApJ, 516, 236 
\bibitem{} Heyrovsk\'y, D., Sasselov, D., \& Loeb, A.\ 2001, ApJ, 543, 406
\bibitem{} Landolt\ 1992, A.U., Astron. J., 104, 340
\bibitem{} Lennon, D.J., Mao, S., Fuhrmann, K., \& Gehren, T.\ 1996, ApJ, 
471, L23
\bibitem{} Paczynski, B. et al.\ 1999, Acta Astron., 49, 319
\bibitem{} Valls-Gabaud, D.\ 1998, MNRAS, 294, 747

\end{thebibliography}
\end{document}